\documentclass[12pt]{article}
\usepackage{geometry}
\usepackage[english]{babel}
\usepackage[utf8x]{inputenc}
\usepackage{url} 

\usepackage{amsmath}
\usepackage{graphicx}
\usepackage{soul}
\usepackage[comma,authoryear,round]{natbib}
\setcitestyle{aysep={}} 
\usepackage{comment}

\usepackage{tikz}
\usetikzlibrary{calc}

\usepackage{fancyhdr} 
\renewcommand{\baselinestretch}{2}

\title{Extending SST Anomaly Forecasts Through Simultaneous Decomposition of Seasonal and PDO Modes}
\author{Rameshan Kallummal}
\date{
	CSIR Fourth Paradigm Institute, NAL Belur campus, Wind Tunnel Road Bangalore - 560 037, Karnataka, INDIA, E-mail: kallummal@gmail.com
	\newline
	\newline
	\today
}

\begin{document}
\maketitle

\begin{abstract}

We present a new approach to forecasting North Pacific Sea Surface Temperatures (SST) by recognizing that interannual variability primarily reflects amplitude changes in four dominant seasonal cycles. Our multivariate linear model simultaneously captures these amplitude-modulated seasonal cycles along with the Pacific Decadal Oscillation (PDO), which naturally emerges as an intrinsic feature of the system rather than a separate phenomenon. Using sixteen-dimensional regression based on four spatially distributed time series per variable, the model delivers unprecedented forecast accuracy for both interannual amplitude modulations and PDO evolution, maintaining skill beyond 36 months—a substantial improvement over current operational and research forecasts, including machine learning methods.  Predictions initialized in 2024 project that the PDO will remain in its negative phase through late 2026, implying reduced likelihood of severe marine heatwaves in the eastern North Pacific during this period. These findings have direct implications for regional climate impacts, including storm tracks, precipitation patterns, and marine ecosystem health. By treating seasonal and interannual variability as coupled rather than independent processes, this framework advances our understanding of North Pacific climate dynamics and provides a powerful tool for stakeholders managing climate-sensitive resources and planning adaptation strategies in regions strongly influenced by North Pacific conditions.
\end{abstract}

\section{Introduction}

Interannual variations in North Pacific SST significantly influence local and global climate, profoundly affecting weather patterns, natural resources, and ecosystems  \citep{mantua1997pacific, deser2004pacific}. Regionally, these fluctuations alter atmospheric circulation, modulating the Aleutian Low's intensity and position, which drives changes in coastal upwelling along western North America and shapes temperature and precipitation across East Asia and the United States  \citep{miller2000interdecadal, fan2017pacific, wu2021poleward}. Globally, they impact ocean-atmosphere teleconnections, affecting rainfall patterns across the pan-North Pacific and extending to eastern North America  \citep{zou2020north, zhang2024enhanced}, tropical cyclone activity  \citep{lee2012multi, yu2016effects, wang2023caused, wang2023pacific,  huang2024combined, dai2025impact}, and marine heatwaves  \citep{joh2017increasing, ren2023pacific}. These dynamics underscore the critical role of North Pacific SST in regulating ecological and climatic processes, making robust seasonal and interannual forecasting an essential goal in climate research.

The North Pacific's primary interannual SST modes - the PDO and Victoria Mode (VM) - emerge from Empirical Orthogonal Function (EOF) analysis of its traditional monthly anomaly, with the PDO as the leading mode and the VM as a secondary dipole  \citep{mantua1997pacific, chiang2004analogous}. The PDO is shaped by local drivers like Aleutian Low variability and Kuroshio-Oyashio Extension dynamics, amplified by midlatitude ocean processes such as wind-driven gyre circulation and Rossby wave propagation  \citep{latif1994causes, newman2016pacific}. Remote forcing from the El Ni\~{n}o-Southern Oscillation (ENSO) via the "atmospheric bridge" further integrates seasonal-to-interannual signals into PDO, while anthropogenic greenhouse gases contribute to long-term trends  \citep{alexander1992midlatitude, zhang1997enso, alexander2002atmospheric}. The VM, influenced by ENSO and the Pacific Meridional Mode, is modulated by stochastic atmospheric forcing  \citep{gershunov1998interdecadal, wang2012seasonality, ding2015influence}. Accurate long-term forecasts of these climate modes are critical for an effective adaptation and resource management, underscoring the urgency of advancing our predictive tools.

The current status of North Pacific SST climate prediction, particularly for PDO, reveals a mix of progress and persistent challenges as outlined here. The constrained predictive skill may stem from the PDO's nature as a complex phenomenon, not a single physical mode, but rather an amalgamation of diverse mechanisms  \citep{newman2016pacific}. Traditional climate models, such as those from Coupled Model Intercomparison Project phases 5 and 6, exhibit low predictive skill for North Pacific SST beyond 12 months  \citep{wen2012seasonal}, largely due to biases in capturing tropical-extratropical teleconnections and ocean dynamics like Rossby wave propagation, leading to uncertainties in long-term forecasts  \citep{ guemas2012identifying, wills2018disentangling,wills2019ocean, zhao2021removing, qin2023pacific}. Linear empirical models, including Linear Inverse Models (LIMs), achieve robust short-term predictions (6-12 months) with correlation scores above 0.5, but their skill diminishes for longer horizons due to nonlinear dynamics and stochastic noise  \citep{schneider2005forcing, alexander2008forecasting}. Nevertheless, the PDO indices employed in some studies  \citep{ alexander2008forecasting, park2013quantitative, huang2020possible} consist of annual data with restricted temporal coverage. The Machine Learning (ML) and Deep Learning (DL) approaches extend PDO forecast lead times up to 12 months by leveraging nonlinear spatiotemporal patterns, outperforming dynamical models  \citep{gordon2021oceanic, qin2023pacific}. However, ML/DL methods face limitations from scarce training data and poor interpretability, while predictions of the VM suffer from inadequate resolution of its dipole structure and high-frequency variability  \citep{bond2003recent}. Overall, while these innovations show promise, systematic biases and incomplete mechanistic representations continue to hinder accurate long-term North Pacific SST predictions  \citep{guemas2012identifying, wang2018twentieth}

Disentangling the combined impacts of external forcings - such as anthropogenic climate change, solar variability, and volcanic eruptions - from the effects of internal drivers, including ocean-atmosphere interactions, stochastic variability, and oceanic gyre dynamics, on SST anomalies presents significant challenges  \citep{an2007influence, nidheesh2017influence}. Such predictors are challenging to estimate accurately due to the intricate nature of climate interactions, overlapping scales of variability, and constraints imposed by limited observational data  \citep{zhao2021removing}. Furthermore, biases in capturing teleconnections introduce additional uncertainty, while non-stationary relationships under a changing climate exacerbate the difficulties in assessment  \citep{ault2018robust}.

We introduce a suite of multivariate linear models built on predictors derived solely from reanalysis data, bypassing explicit dependence on remote or internal forcing predictors. Our findings show that these reanalysis-based predictors markedly improve statistical model performance in forecasting interannual variability, despite not directly tackling inherent predictive challenges.

\section{Methods and Data}

\noindent{\bf Observational Data}

We analyze monthly data from 1948 to 2025, including SST from NOAA Extended Reconstructed SST version 5 \citep{huang2017extended} and surface winds (zonal and meridional components) plus surface pressure from NCEP/NCAR Reanalysis \citep{kalnay2018ncep}. For brevity, we denote SST as `S', zonal wind as `U', meridional wind as `V', and surface pressure as `P'. We deliberately selected observational products with minimal model-based gap-filling to avoid uncertainties from model-generated estimates. Variables like ocean heat content and thermocline depth, which rely heavily on model simulations, were excluded to maintain an empirically grounded prediction framework based solely on observed data.

\noindent{\bf Extraction of Seasonal Cycles}

Following the framework established in \citet{kallummal2021deciphering, kallummal2023sea, kallummal2024decoding}, we extract four dominant seasonal modes from each variable across the North Pacific (20$^o$N to 60$^o$N; 100$^o$E to 90$^o$W) using Singular Value Decomposition (SVD) applied directly to raw, uncentered data. This approach identifies statistically significant patterns without imposing predetermined assumptions about seasonal structure.

SVD factorizes the data matrix $\mathbf{\Psi}$ (with $T$ time samples in columns and $G$ spatial samples in rows) into three components: \[\mathbf{\Psi} = \mathbf{L}\mathbf{\Sigma} \mathbf{R}^{{\tau}},\] where $\mathbf{L}$ and $\mathbf{R}$ are orthogonal matrices and $\mathbf{\Sigma}$ contains singular values in descending order. The superscript $\tau$ denotes matrix transpose. Because we apply SVD to the complete dataset without monthly stratification, the first four columns of $\mathbf{R}$ represent spatial patterns for seasonal modes 1 through 4, while corresponding columns of $\mathbf{L}$ capture their temporal evolution.

Each SVD mode's contribution at time $\mathbf{t}$ and grid point $\mathbf{g}$ is reconstructed as: \[\mathbf{\large\psi\mathrm{{\bf i_{t,g}}}} = \mathbf{{l}_{t,i}} \times \mathbf{\sigma_{i}} \times \mathbf{r_{g,i}}, \mathrm{where\: i=1,2,3,4.}\]

We denote the four leading modes for SST as S1, S2, S3, and S4, with analogous notation for other variables. The suffix 1, 2, 3, and 4 indicate the four leading SVD modes. Thus, the first SVD modes of the four variables are S1, U1, V1, and P1, with subsequent modes denoted by replacing 1 with 2, 3, or 4. Supplementary figures S2 through S5 demonstrate the dynamically coupled evolution of these long-term mean seasonal cycles.

By extracting seasonal modes independently for each variable, we avoid spurious relationships that can arise from multivariate statistical methods. The dynamic coupling among variables (Figs. S2 to S5) emerges naturally, reflecting physically meaningful co-variations. Additionally, analyzing uncentered data ensures we capture spatio-temporal variations without reference to an arbitrary time-invariant background state. The singular values $\mathbf{\sigma}$ from raw data correspond to root mean square values rather than variances, meaning our ``seasonal cycles'' and ``modes'' represent structures capturing substantial portions of mean squared  values. This methodology provides robust analysis and clearer insight into coupled seasonal evolution across all four variables.

\noindent{\bf Interannual Amplitude Modulations}

We calculate interannual amplitude modulations (Figures S1 and 1G), referred to as anomalies, by removing each mode's long-term seasonal cycle (black lines in Figures. S1 and 1G). This procedure applies uniformly across all modes and variables. We denote these anomalies by prefixing `a' to variable names: aS1, aS2, aS3, and aS4 for SST anomalies, with consistent notation for other variables.

Previous investigations of low-frequency variability in tropical Pacific SST \citep{kallummal2023sea, kallummal2024decoding} identified aS1 as representing secular warming, while the third and fourth modes correspond to central and eastern Pacific ENSO, respectively. The second coupled seasonal cycle (S2, U2, V2, P2) was found to stabilize the climate system by moderating external forcings \citep{kallummal2024decoding}, as evidenced by its relatively small interannual variations. This stabilizing characteristic, manifested through small interannual variations, is similarly evident in the NP basin seasonal modes examined here (second column, Fig. S1).

\noindent{\bf Regional Indices}

Given that North Pacific SST anomalies—particularly the PDO—can influence regional seasonal predictability, we target high-variance regions within each variable's four seasonal mode anomalies (see {\bf Interannual Amplitude Modulations} in section 2 and Fig. 2A and Table 1). Four area-averaged indices per variable are derived from maximum-variance locations distributed across the North Pacific basin (Fig. 2A). For SST (i.e., S) , these indices span: IaS1 (152°E–156°E, 28°N–32°N), IaS2 (146°E–150°E, 48°N–52°N), IaS3 (126°E–130°E, 28°N–32°N), and IaS4 (140°W–144°W, 48°N–52°N), with corresponding time series presented in Figure 2B (top panel).

To suppress subseasonal noise, we smooth each index with a 3-month running window---a standard technique for enhancing climate forecast skill by emphasizing low-frequency variability \citep{choudhury2015predictability, newman2016pacific}. Since PDO dynamics operate on timescales exceeding three months, this smoothing isolates persistent interannual signals that dominate NP variability, thereby improving forecast performance. To maintain causality, smoothed values are assigned to the final date in each 3-month window, consistent with operational practices at institutions like the International Research Institute for Climate and Society.

The four leading SVD modes of each variable are mutually orthogonal, ensuring statistical independence. However, their anomalies (i.e., aS, aU, aV, and aP) exhibit temporal correlations even at zero lag, indicating underlying physical processes connecting these variables in interannual climate variability. The effectiveness of our regression-based prediction thus depends on cross-correlations among all 16 precursors: IaS1–IaS4 (SST), IaU1–IaU4 (zonal wind), IaV1–IaV4 (meridional wind), and IaP1–IaP4 (SLP).

\noindent{\bf Linear Regression Model}

Following \citet{halide2008complicated}, who demonstrated advantages of simple linear regression for climate forecasting over complex climate models and machine learning approaches \citep{jin2008current, ham2019deep, zhou2023self, l2020enso, barnston2012skill, gonzalez2016long}, we employ a multivariate linear model. Linear approaches offer several benefits: computationally efficient long-term forecasts, transparent and interpretable results showing individual variables' influences, and minimal training data requirements---particularly valuable when data is limited.

Our model estimates future precursor values based on past values of all 16 precursors:

\begin{equation}
	x_j(t) = \sum_{i=1}^{16} p_i(l) x_i(t-l)  + e_j(t)
\end{equation}

where $p_i(l)$ represents the regression coefficient at time lag $l$, $x_i$ belongs to the ordered set \{IaS1, IaS2, IaS3, IaS4, IaU1, IaU2, IaU3, IaU4, IaV1, IaV2, IaV3, IaV4, IaP1, IaP2, IaP3, IaP4\}, $x_j$ represents one of the ordered set \{IaS1, IaS2, IaS3, IaS4\}, and $e_j(t)$ is the error term.

We estimate regression coefficients $p_i(l)$ by minimizing error using Python's least-squares fitting routine ({\it e.g.,} numpy.linalg.lstsq), focusing predictive skill evaluation on the four leading SST modes: $j = $1 to 4.

The training period spans 53 years (1948 to 2000), providing robust temporal sampling for accurate coefficient estimation. Verification covers January 2001 to December 2023, with the final year (January 2024 to November 2025) reserved for out-of-sample forecast evaluation (Fig. 5).

\noindent{\bf Forecast Generation and Verification}

Spatially remote and temporally lagged anomalies both influence PDO evolution and other interannual signals \citep{newman2003enso, newman2016pacific}. Our multivariate regression model explicitly accounts for remote influences by incorporating precursors distributed across the North Pacific (Figure 2A and Table 1), which are derived from multiple SVD modes of all four variables.

We generate predictions using 36 regression equations, each corresponding to a specific time lag $l$ between predictors and predictands, capturing lagged influences. The predicted value $y_j^l(k)$ for the $j^{th}$ SST index at future time $k$ derives from past predictor values $x_i$ at time $(k-l)$:

\begin{equation}
	y_j^l(k) = \sum_{i=1}^{16} p_i(l) x_i(k-l) ,
\end{equation} 

where $k$ = $l$, $l$ + 1, $l$ + 2, ..., $N_t$ + 36, $j = $1 to 4 ({\it i.e.,} only S), and $l$ = 1, 2, ..., 36. $N_t$ indicates the time-index of the final verification date ({\it i.e.,} December 2023). This enables out-of-sample predictions extending 36 months beyond the final verification date.

To account for lagged influences robustly, we average predictions across multiple lead times ($l$). For greater statistical robustness, only median values of these predictions are used, as this approach reduces outlier impacts and yields more reliable forecasts. 

\noindent{\bf Skill Metrics}

This study evaluates forecast skills using the Pearson Correlation Coefficient (PCC) and Root Mean Square Error (RMSE). Estimations of PCC and RMSE are performed over values of linear regression hindcast indices ($y$) and precursor indices ($x$) during the verification period using the following formulae:

\noindent $PCC = \frac{cov(y,x)}{s_xs_y}$ where $cov()$ indicates the covariance estimation procedure; $s_y$ and $s_x$ denote standard deviations of $y$ and $x$, respectively.

\noindent $RMSE = \sqrt{\frac{1}{N_t} \sum_{t=1}^{N_t} (x_t-y_t)(x_t-y_t)}$. For out-sample forecasts (Fig. 5), $t =1$ corresponds to January 2023 and $t = N_t$ to December 2023.

\noindent{\bf Spatial Field Reconstruction}

The 16 precursors (Fig. 2B) capture temporal evolution in high-variance regions distributed across the North Pacific (Figure 2A and Table 1), providing a comprehensive representation of PDO complexity. To assess the model's ability to reproduce total interannual SST variability, we compare observed spatiotemporal anomalies (aS1 + aS2 + aS3 + aS4) with the corresponding regression-based predictions.

\noindent Spatial field reconstruction from predicted indices proceeds as follows:

\begin{enumerate}
	\item For each grid point in the aS1 field, we estimate regression coefficients relating the local aS1 time series to the IaS1 index.
	\item We multiply spatial regression coefficient patterns by the IaS1 index to generate regressed spatio-temporal anomalies.
	\item This procedure repeats for aS2, aS3, and aS4 using their respective indices IaS2, IaS3, and IaS4.
\end{enumerate}

For both verification and out-of-sample periods, we multiply spatial regression patterns by hindcast precursors (verification period) or forecast precursors (out-of-sample period) to generate complete spatio-temporal anomaly fields.

\noindent{\bf Statistical Significance}

Calculated t-values for both regression and correlation coefficients exceed critical t-values at a 98.5\% confidence level, confirming statistical significance of all relationships.
\section{Results}

\noindent{\bf Four Amplitude Modulated SCs and SC4 alias PDO}

The canonical PDO and VM indices are derived from eigen modes of monthly Pacific SST anomalies north of 20$^o$, calculated as deviations from the mean seasonal cycle at each spatial location. This method hinges on a critical assumption that the primary responses of the ocean-atmosphere system can be effectively represented by a single seasonal cycle with constant amplitude. However, this assumption lacks dynamical grounding, as it fails to account for the possibility of multiple seasonal cycles, each characterized by distinct, non-sinusoidal spatiotemporal patterns and amplitudes that fluctuate interannually  \citep{kallummal2021deciphering, kallummal2022seasonal, kallummal2022drivers, kallummal2023sea, kallummal2024decoding}. Comparable to the externally forced fundamental modes of a complex dynamical system, these seasonal cycles emerge predominantly from interactions among land, ocean, and atmosphere, driven by periodic solar radiation and amplified through nonlinear processes.

Traditionally, seasonal cycles (SCs) of North Pacific SST and the dominant low-frequency modes (PDO and VM) have been regarded as distinct phenomena that interact with one another. Consequently, low-frequency modes are typically derived from anomalies relative to a mean seasonal cycle \citep{newman2007interannual}. However, this conventional approach assumes their independence a priori, rather than allowing them to emerge naturally from spatiotemporal decomposition, as demonstrated for ENSO by \citet{kallummal2021deciphering, kallummal2024decoding} and for PDO by \cite{kallummal2022seasonal}. Moreover, EOF analyses of traditional anomalies often suffer from mode mixing, where signals from ENSO, PDO, and global warming become conflated \citep{wills2018disentangling}. To address this issue, researchers commonly subtract the time-dependent basin-mean anomaly from the traditional SST anomaly prior to EOF analysis \citep{newman2016pacific, deser2000teleconnectivity}, or employ additional techniques to disentangle the overlapping signals \citep{xu2022increase}.

Strikingly, a low-frequency mode with minimal seasonality (Fig. 1G) emerged as the fourth SVD mode of S, mirroring the canonical PDO in spatiotemporal patterns (Figs. 1A-F). Thus, we interchangeably denote the fourth SVD mode as S4 or PDO. This concurrent extraction of seasonal ({\it i.e.}, S1, S2, and S3) and low-frequency ({\it i.e.}, S4 alias PDO) modes from SST is unprecedented. Notably, S4 exhibits no evident secular warming trend. The latter is captured by S1 and S2, thereby eliminating the need for additional techniques to disentangle PDO and secular warming signals. However, the secular trend is not evident in the S1 and S2 precursors shown in Fig. 2B (top panel), as all precursors exhibiting secular trends have been detrended prior to linear regression. Conversely, in the tropical Pacific \citep{kallummal2021deciphering, kallummal2024decoding} and Indian Ocean \citep{kallummal2022seasonal}, such trends are confined to S1 only.

\noindent{\bf Empirical Model Verifications}

The all-season Pearson Correlation Coefficients (PCC) for IaS1, IaS2, IaS3, and PDO indices (black line, Fig. 3A) exceeds 0.5 even when the lead times are beyond 36 months, with corresponding Root Mean Square Error (RMSE) values (red-dash line, Fig. 3A) consistently below 0.5. Predicted time series (colored lines, Fig. 3B) closely replicate key features of observed time series (black lines, Fig. 3B) across all lead times. Our linear regression models exhibit superior skill in capturing NP variability compared to operational forecast models  \citep{hu2014prediction, wen2012seasonal}. Notably, prior studies could assessed only PDO mode's performance skills, as the three seasonal modes' low-frequency amplitude modulations were not resolvable from the traditional anomalies. Furthermore, our model outperforms ML/DL approaches \citep{qin2022deep, gordon2021oceanic} in PCC values and sustains RMSE below 0.5 across all four IaS and PDO indices. This precision is striking, as earlier studies typically yielded RMSE values exceeding 0.5 for lead times beyond 9 months.

Although improved all-season forecasting skill, demonstrated by PCC and RMSE across all time samples, indicates enhanced predictive capacity, it does not guarantee accurate prediction of significant interannual events’ spatio-temporal evolution.
To investigate this, we compared the observed spatial evolution of total interannual variability ({\it i.e.,} aS1 + aS2 + aS3 + aS4; Fig. 4B) with the summed forecasts (Fig. 4A) from April 2017 to July 2024. 
Both observed and predicted patterns for warm and cold events showed strong spatio-temporal alignment. This sustained congruence over seven years bolsters confidence in our forecasting approach. Previous studies, as pointed out earlier, have failed to resolve the spatial distribution of year-to-year variations in S1, S2, S3, S4 modes.

These findings significantly enhance our ability to forecast NP's low-frequency dynamics, overcoming limitations of prior models that struggled to capture seasonal mode interactions alongside PDO evolution \citep{schneider2005forcing, wills2019ocean}. Earlier efforts, such as those using operational forecasts, achieved modest skill at shorter lead times but faltered beyond 6 months \citep{hu2014prediction}, while ML/DL approaches, though innovative, often missed fine-scale spatial patterns \citep{qin2023pacific}. By integrating three interannual amplitude modulations exhibited by three seasonal cycles ({\it i.e.,} aS1, aS2, and aS3) with PDO ({\it i.e.,} aS4 ), our model accurately tracks anomaly progression, potentially offering vital insights for marine ecosystem forecasting and regional climate resilience.

\noindent{\bf Out-of-sample Predictions}

Having validated the reliability of our 16-dimensional linear regression model, the critical challenge in climate prediction lies in producing accurate out-of-sample forecasts. However, the definitive evaluation of such forecasts requires comparing predictions with future observations as events unfold. For this purpose, the forecasts were initiated during 2023 for the out-of-sample predictions from 2024 to 2027 (Fig. 5).

The spatio-temporal evolution of the total predicted anomaly (aS1 + aS2 + aS3 + aS5 + PDO) from December 2024 to October 2027 indicates a persistent negative PDO phase throughout 2025, characterized by negative SST anomalies in the eastern NP and positive anomalies in the western NP (Fig. 5B). This projection is consistent with recent observations showing that the PDO has remained in a strongly negative phase since approximately 2020. Data from NOAA's National Centers for Environmental Information (NCEI; \url{https://www.ncei.noaa.gov/access/monitoring/pdo/}) confirm that PDO values have remained consistently negative throughout 2024 and 2025.

\noindent {\bf Sensitivity to Training Period and Model Robustness}

To assess the robustness of our results, we conducted additional analyses using a regression model trained on data spanning 1948 to 1979 (Fig. 5B). The predictions generated by this alternative model are comparable to those obtained using the original model trained on 1948–2000 data (Fig. 5A). The comparison of Figs. 5A and 5B with the spatio-temporal evolution of corresponding SST anomalies observed from January through October 2025 (Fig. 5C) demonstrates remarkable predictive accuracy. The qualitative agreement between predictions from the two models further substantiates the robustness of our results. To further validate our approach, we evaluated the skill of ten regression models trained across distinct time periods, which confirms the robustness of our results and excludes chance effects (Fig. 6). The linear regression models exhibit consistent performance, producing comparable skill levels regardless of the training period employed. This underscores the critical role of interactions among the twenty indices in achieving accurate interannual forecasts across different decades

\section{Implications of a Negative PDO State on Weather}
A negative PDO phase through 2025, driven by a stronger Aleutian Low and enhanced upwelling in eastern NP waters  \citep{mantua1997pacific, miller2000interdecadal}, will affect: 1) Cyclones: Fewer intense typhoons in the western NP occur during a negative PDO phase, linked to cooler SSTs and a stronger subtropical high, with modulation by large-scale circulation changes  \citep{choi2010change,qin2024relationship}. 2) Rainfall: Less precipitation in southern China and Japan from a weaker monsoon  \citep{gershunov1998interdecadal}, drought risk in the southwestern U.S.  \citep{zou2020north}, and more rain in the central NP  \citep{mantua1997pacific}. 3) Marine Heatwaves: Reduced frequency along the eastern NP as cooler SSTs limit warm anomalies  \citep{joh2017increasing}. 4) Marine Life: Benefits for species like salmon   \citep{mantua1997pacific, newman2016pacific}, with upwelling boosting fishery productivity  \citep{miller2000interdecadal}. 5) Monsoons: Weaker East Asian and North American monsoons, reducing rainfall  \citep{gershunov1998interdecadal}.

The PDO modulates the impacts of ENSO through its phase. When the PDO and ENSO are in the same phase (e.g., negative PDO and La Niña), their climatic impacts are often magnified. La Niña events during a negative PDO phase result in approximately 20\% more global land area experiencing abnormal wetness or flooding compared to warm PDO phases. When they are in opposite phases (e.g., negative PDO and El Niño), the PDO acts to offset or dampen the typical ENSO teleconnections. El Niño impacts over North America, such as warm surface anomalies in Alaska, are diminished during a negative PDO.

These changes underscore the broader climatic influence of the PDO, with teleconnections extending to precipitation and cyclone activity across the pan-North Pacific and eastern North America  \citep{park2017interannual}. Collectively, these implications - drawn from a forecast extending beyond 36 months with high skill (Fig. 5a) - highlight the negative PDO’s role in shaping weather extremes and ecological responses. This predictive insight, surpassing the limited lead times of prior models  \citep{wen2012seasonal,qin2023pacific}, enhances preparedness for climate-driven impacts on agriculture, water resources, and marine biodiversity through 2025.

\section{Conclusion}
This study introduces a transformative approach to NP SST forecasting by integrating amplitude-modulated SCs with PDO in a multivariate linear regression framework. Unlike prior efforts that treated seasonal and low-frequency variability as distinct phenomena, our method simultaneously extracts four dominant SCs alongside the PDO as the fourth leading SVD mode, without any priori assumptions and processing of data. This decomposition reveals the PDO as an emergent feature of NP dynamics, distinct from traditional EOF-based anomaly analyses that often conflate signals  \citep{wills2018disentangling, newman2016pacific}. The resulting sixteen-dimensional model achieves unprecedented predictive skill, with all-season PCC exceeding 0.5 and RMSE below 0.4 for lead times beyond 36 months (Fig. 3a), far surpassing the capabilities of earlier models, for PDO.

Previous attempts at PDO forecasting faced significant limitations. In contrast, our model not only outperforms these benchmarks - evidenced by superior PCC and RMSE across all IaS and PDO indices - but also resolves the spatio-temporal evolution of total anomalies (aS1 + aS2 + aS3 + aS5 + PDO) with high fidelity over seven years (Fig. 4). Out-of-sample forecasts from 2024 to 2027 further predict a sustained negative PDO phase through 2025 (Fig. 5B), a capability unattainable in prior studies that focused solely on PDO without seasonal mode interactions  \citep{hu2014prediction,wills2019ocean}.

The extraction of amplitude-modulated SCs is pivotal to this advance. By accounting for time-varying amplitudes within the seasonal framework, our approach  \citep{kallummal2022seasonal, kallummal2021deciphering, kallummal2022drivers} captures critical interannual signals overlooked by traditional methods, which assume a static seasonal cycle. This innovation enables the model to disentangle overlapping variability - such as ENSO, PDO, and global trends - offering a clearer mechanistic picture of NP SST dynamics. The robustness of our test-forecasts, validated across training periods (Figs. 5 and 6), underscores the stability of these interactions across decades, enhancing confidence in long-term predictions.

Knowing the importance of sixteen key predictors, we suggest starting a regular monitoring program in the high-variance areas our study identified (Fig. 2a). This enhanced data collection will facilitate a more detailed understanding of the links between atmospheric variability and oceanic processes, paving the way for improved seasonal climate prediction. Such a targeted, data-driven approach underscores the potential for significant advancements in this field.

In conclusion, this study redefines North Pacific SST forecasting by bridging seasonal and decadal scales with exceptional accuracy and detail. Its ability to predict anomaly evolution over 36 months, coupled with insights into negative PDO persistence, provides a powerful tool for climate resilience, resource management, and ecosystem planning. These advancements address longstanding challenges in climate prediction  \citep{newman2016pacific}, setting a new standard for statistical modeling in atmospheric and oceanic sciences.

\vspace{1 cm}
\noindent {\large \bf Acknowledgements:} {Author acknowledges the computational facilities at CSIR-4PI. The visualization tools used are GrADS (http://cola.gmu.edu/grads/grads.php) and Matplotlib (https://matplotlib.org/). This study and author have received no funding. 

\vspace{1cm}
\noindent {\large \bf Data Availability} {Data used is available from \protect{\url{https://psl.noaa.gov/data/gridded/data.noaa.ersst.v4.html}} \protect{\cite{huang2017extended}} and \protect{\url{https://psl.noaa.gov/data/gridded/data.ncep.reanalysis.surface.html}} \protect{\cite{kalnay2018ncep}}.}

%



\typeout{}
\bibliographystyle{spbasic}      
\bibliography{npac_cleaned}   
\newpage
\renewcommand{\baselinestretch}{2}

\begin{figure}
     \begin{center}
        \includegraphics[scale=1.0]{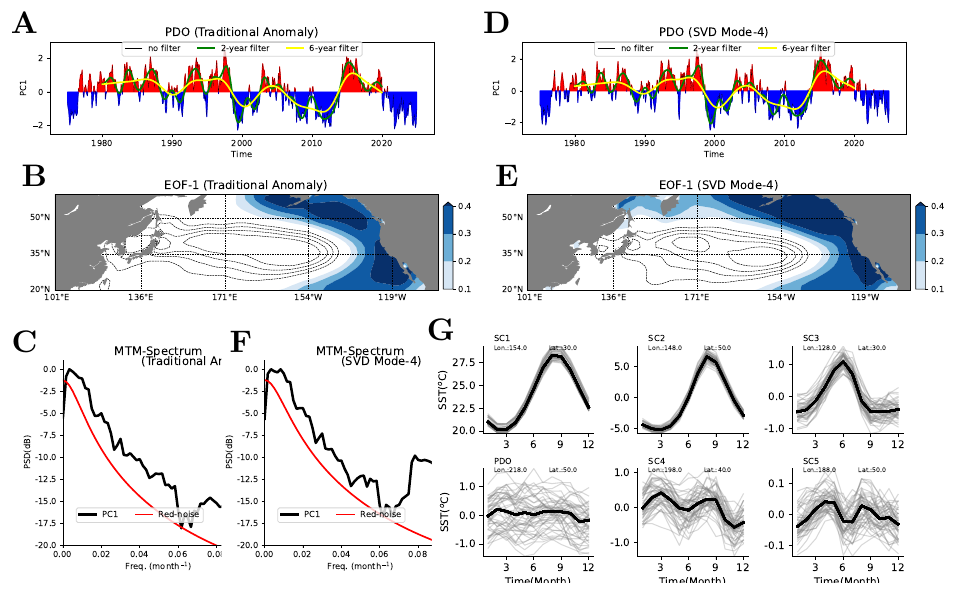}
     \end{center}

        \caption{ PDO characteristics including area-mean time series, spatial pattern, and power spectrum: {\bf A, B, C} first eigenmodes of the traditional anomaly; {\bf D, E, F} aS4. {\bf G} Area-averaged interannual amplitude modulations for leading SST SVD modes within their high-variance regions. Each panel displays 49 twelve-month segments spanning January to December, with the midpoint coordinates of the high-variance region indicated. Thick black lines denote long-term means.}
\end{figure}

\begin{figure}
     \begin{center}
        \includegraphics[scale=1.5]{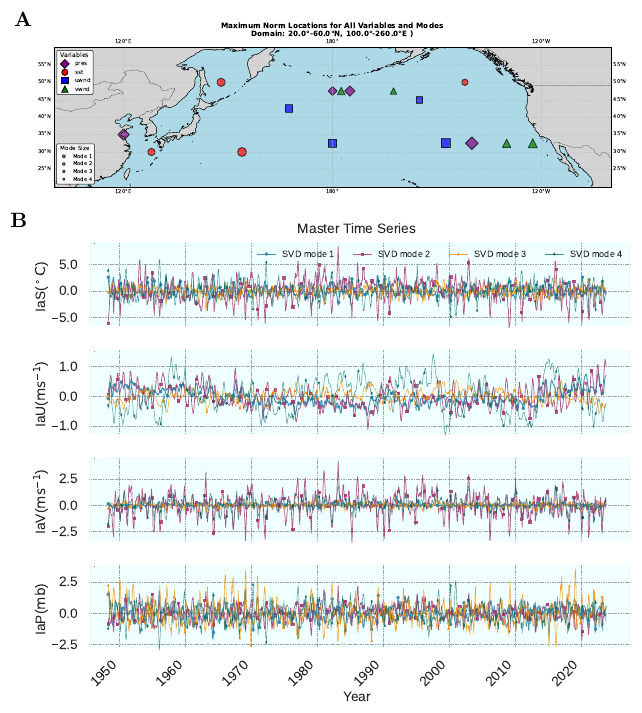}
     \end{center}
        \caption{Locations of high-variance regions and their area-averaged time series. {\bf A} Targeted observation regions (symbol sizes decrease from SVD modes 1 to 5, with four high-variance regions per variable; coordinates in Table 1). {\bf B} Area-mean time series depict deviations from long-term seasonal cycles. Detrended S1 and S2 (top panel), which originally captured SST secular warming trends.}
\end{figure}

\begin{figure}

	\renewcommand{\figurename}{Table} 
	\renewcommand{\thefigure}{1}      
	\includegraphics[width=6.5in]{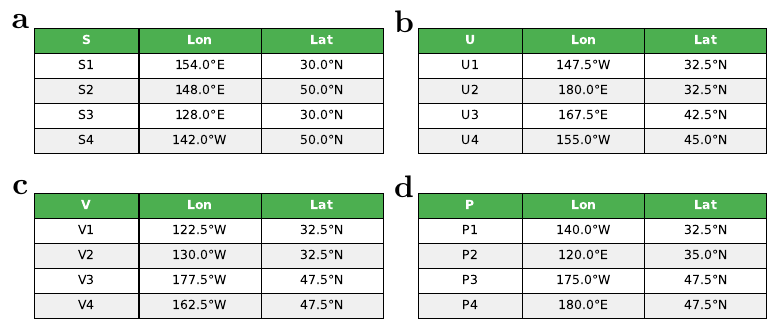}
	\caption{Coordinates of the high-variance regions: (a) SST, (b) U, (c) V, and (d) P}
\end{figure}

\begin{figure}
	\renewcommand{\figurename}{Figure} 
	\renewcommand{\thefigure}{3}      
     \begin{center}
        \includegraphics[scale=1]{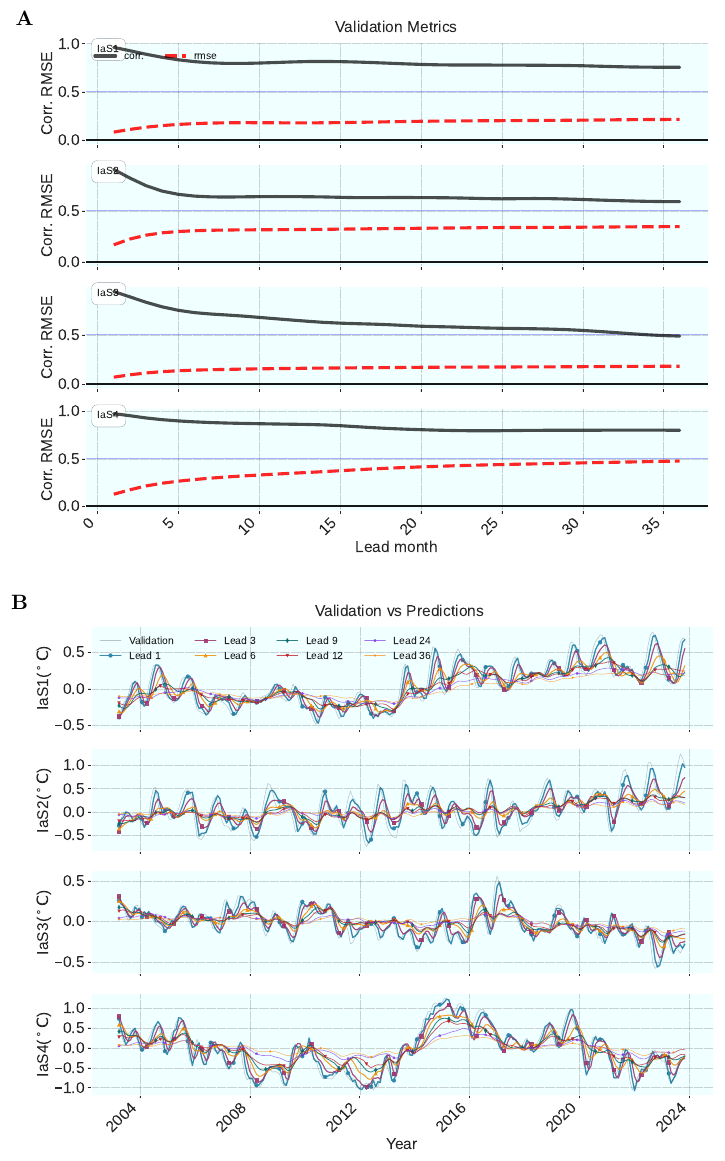}
     \end{center}
        \caption{All-season predictive skill and forecasted time series in the 2000–2023 verification period. {\bf A} Correlation (black lines) and MSE (red lines) versus lead time. {\bf B} Median anomalies (colored lines) across lead times, compared to the reanalysis data during validation period (black line).}
\end{figure}

\begin{figure}
	\renewcommand{\figurename}{Figure} 
	\renewcommand{\thefigure}{4}      
     \begin{center}
        \includegraphics[scale=1.35]{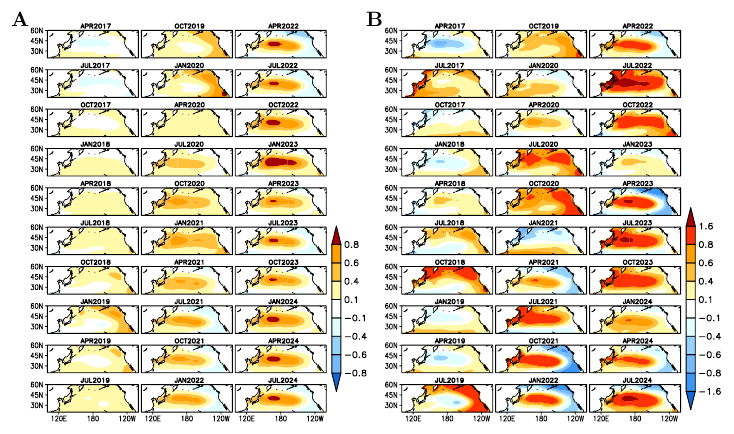}
     \end{center}
        \caption{SST spatio-temporal anomalies, predictions during 2017-2024 period. {\bf A} Median anomaly patterns (aS1 + aS2 + aS3 + aS4 + aS5). {\bf B} Corresponding reanalysis anomalies.}
\end{figure}

\begin{figure}
	\renewcommand{\figurename}{Figure} 
	\renewcommand{\thefigure}{5}      
     \begin{center}
        \includegraphics[scale=1.3]{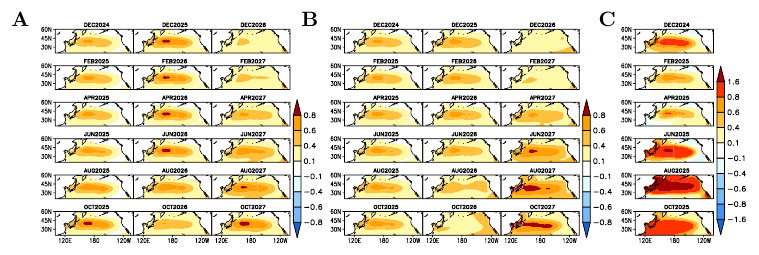}
     \end{center}
        \caption{Out-of-sample forecasts of SST anomalies initiated from the final months of 2024. {\bf A}  and {\bf B} Predicted patterns of total anomaly ({\it i.e.,} aS1 + aS2 + aS3 + aS4 + aS5) from December 2024 to October 2027. {\bf C} Corresponding anomaly estimates from observations available through October 2025. The linear regression models were estimated using data from 1948 to 2000 for {\bf A} and from 1948 to 1979 for {\bf B}.}"
  
\end{figure}

\begin{figure}
	\renewcommand{\figurename}{Figure} 
	\renewcommand{\thefigure}{6}
	\begin{center}
     \begin{tikzpicture}

		\node[inner sep=0pt] (P1) at (0,11) {\noindent\includegraphics[scale=0.27,angle=0]{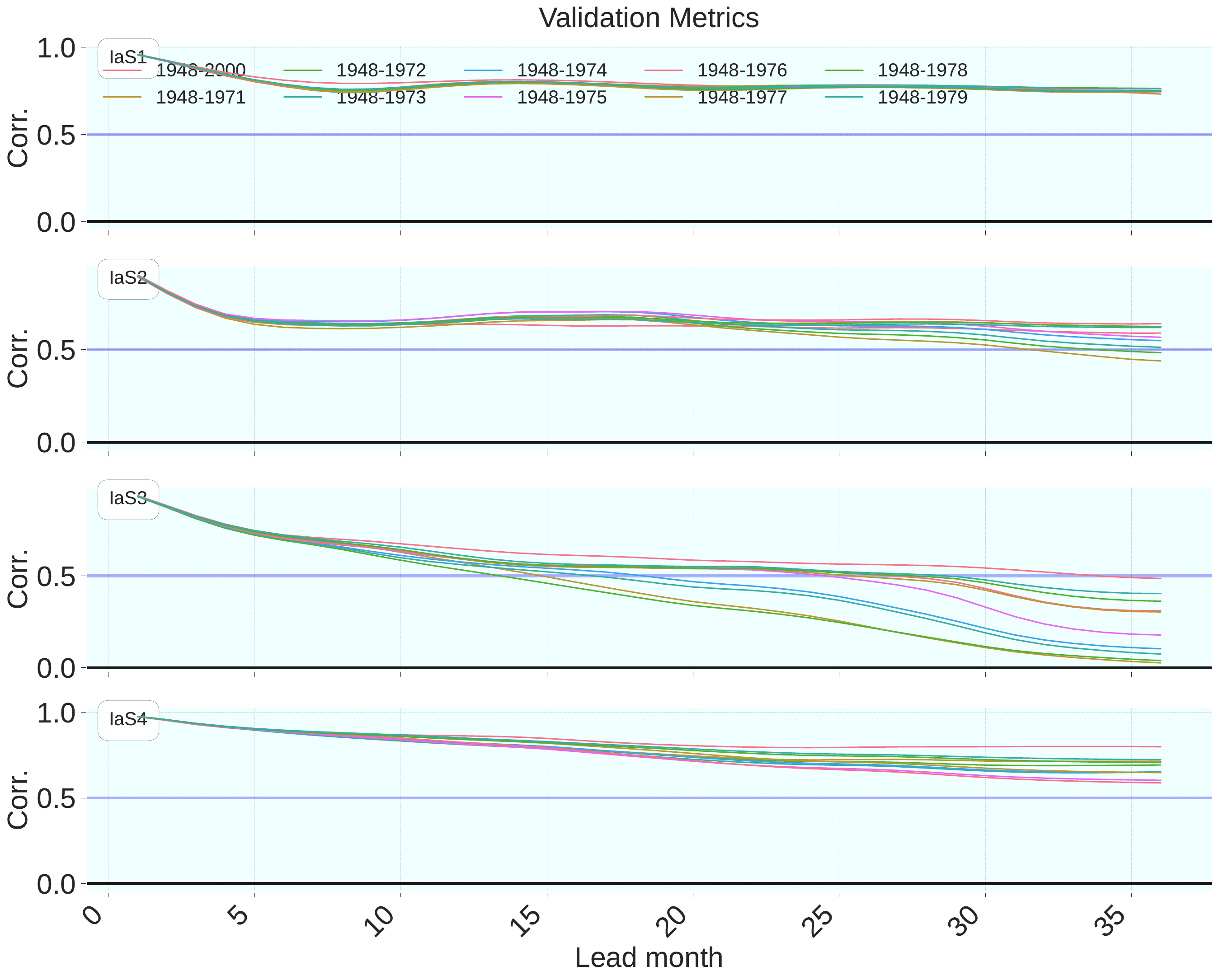}}; \node[black, ultra thick, scale=1.5] at ($(P1.north west)!-0.02!(P1.north east)$) { \bf A};

		\node[inner sep=0pt] (P1) at (0,1) {\noindent\includegraphics[scale=0.27,angle=0]{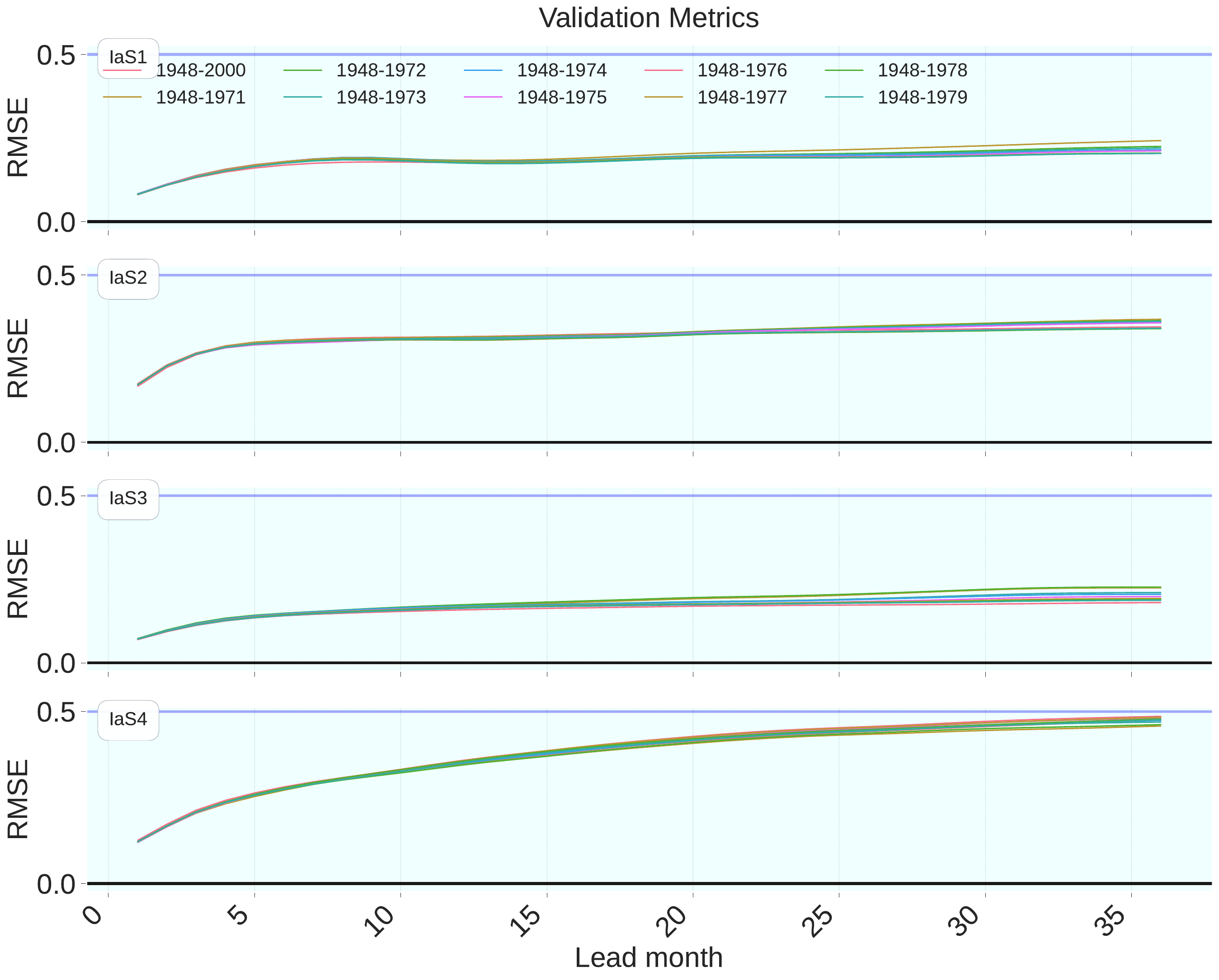}}; \node[black, ultra thick, scale=1.5] at ($(P1.north west)!-0.02!(P1.north east)$) { \bf B};
		
	\end{tikzpicture}
    \end{center}
	\caption{Correlation skills across all seasons during the verification period. (A) PCC and (B) RMSE as functions of lead time. Colored lines represent the performance of 10 additional regression models trained on data from the following periods: 1970-2000, 1971-2000, 1972-2000, 1973-2000, 1974-2000, 1975-2000, 1976-2000, 1977-2000, 1978-2000, and 1979-2000.}
	
\end{figure}

\clearpage
\vspace*{\fill}
\begin{center}
	{\Large \bf {Supplementary figures for "Extending SST Anomaly Forecasts Through Simultaneous Decomposition of Seasonal and PDO Modes"}} {\\ Rameshan Kallummal \\ CSIR Fourth Paradigm Institute, NAL Belur campus, Wind Tunnel Road Bangalore - 560 037, Karnataka, INDIA \quad kallummal@gmail.com}
\end{center}
\vspace*{\fill}
\clearpage

\begin{figure}
        \includegraphics[width=6.5in]{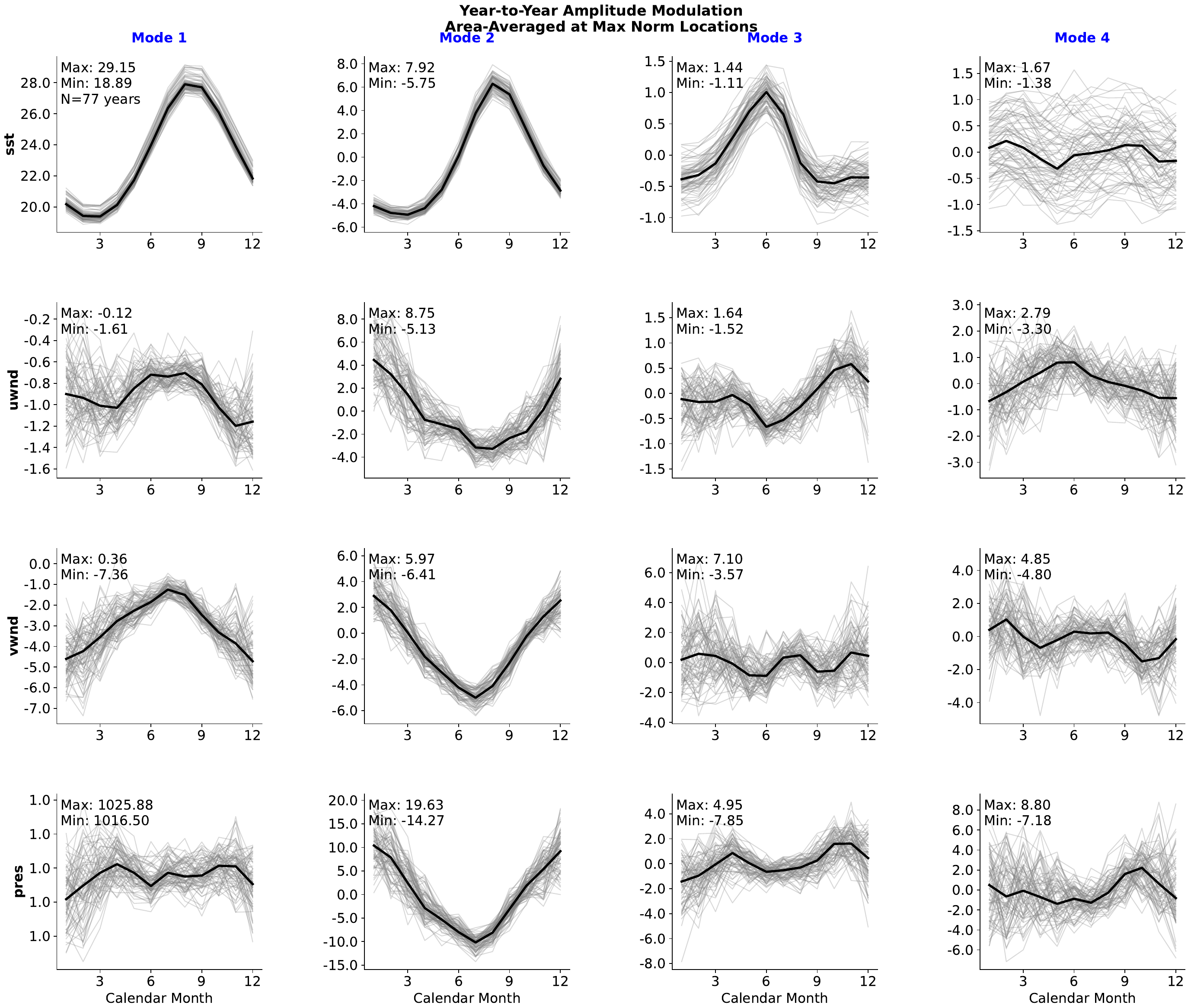}
        \renewcommand{\figurename}{Figure}
        \renewcommand{\thefigure}{S1}
        \caption{Time series of the four leading seasonal cycles (columns) for four variables (rows).
                SC denotes the seasonal cycle. Each panel contains fifty-three seasonal cycles (from January to December). Mid-point of the high-variance region is indicated in each panel. Thick black lines represent long-term mean seasonal cycles}
\end{figure}

\pagebreak[8]

\clearpage

\begin{figure}
        \includegraphics[width=\textwidth]{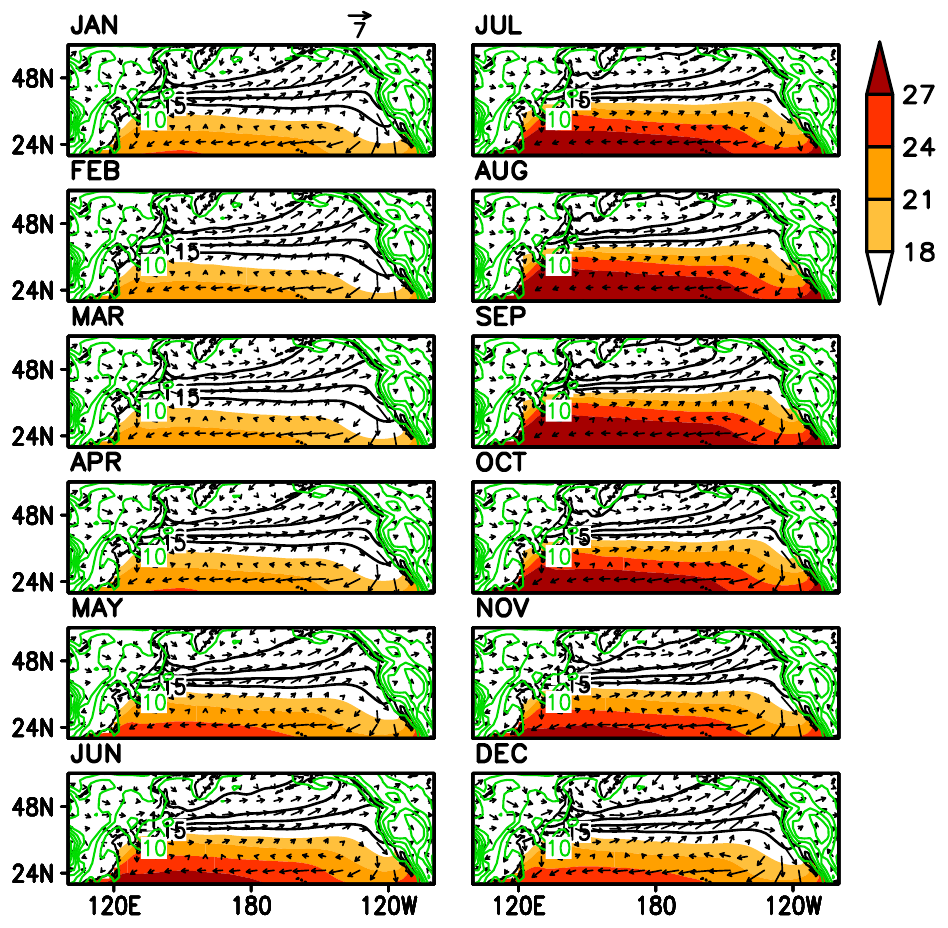}
        \renewcommand{\figurename}{Figure}
        \renewcommand{\thefigure}{S2}
        \caption{Long-term mean seasonal cycles of first SVD modes. SST (shaded; o C), wind (vectors; m sec -1 ) and SP (green contours, mbar)}
\end{figure}

\begin{figure}
        \includegraphics[width=\textwidth]{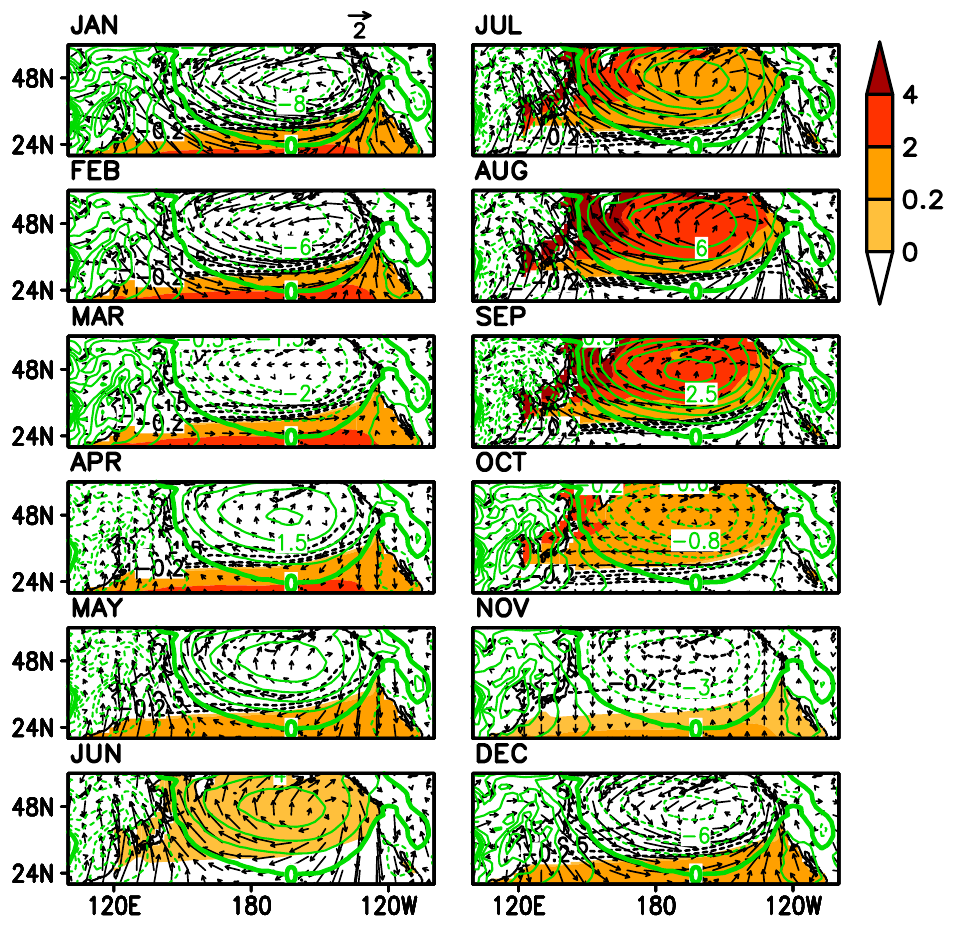}
        \renewcommand{\figurename}{Figure}
        \renewcommand{\thefigure}{S3}
        \caption{Long-term mean seasonal cycles of second SVD modes. SST (shaded; o C), wind (vectors; m sec -1 ) and SP (green contours, mbar)}
\end{figure}

\begin{figure}
        \includegraphics[width=\textwidth]{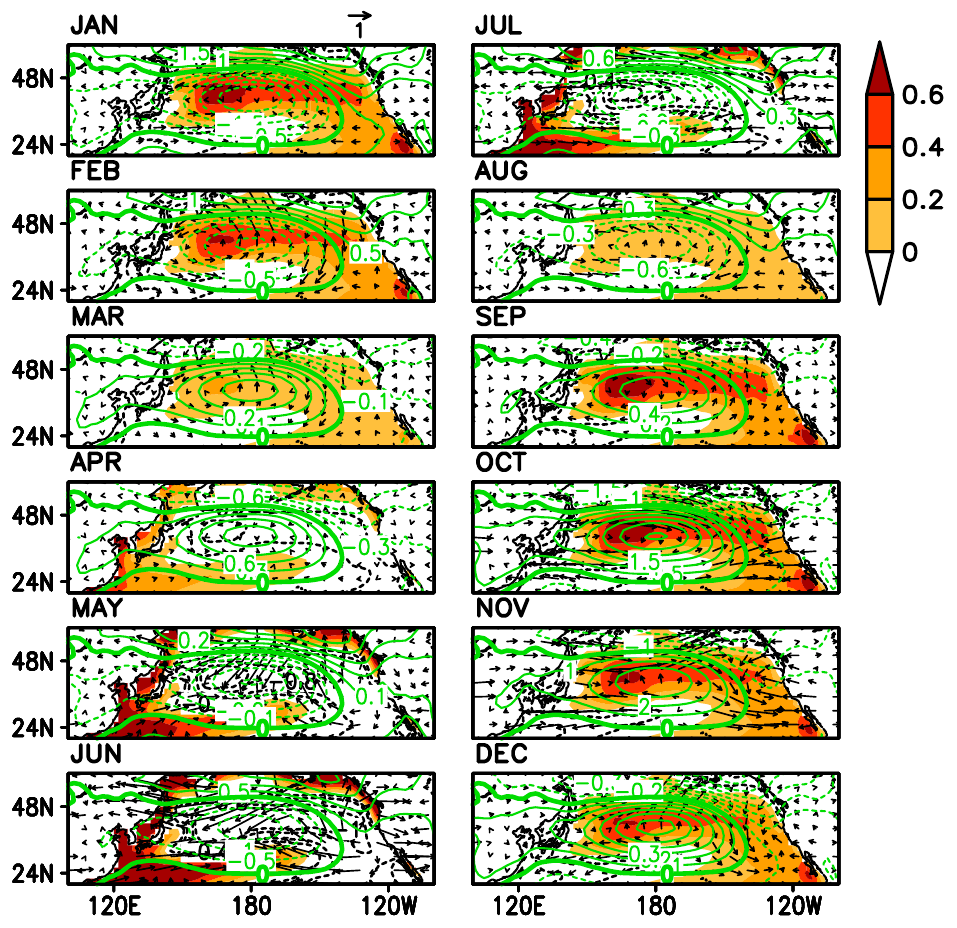}
        \renewcommand{\figurename}{Figure}
        \renewcommand{\thefigure}{S4}
        \caption{Long-term mean seasonal cycles of thrid SVD modes. SST (shaded; o C), wind (vectors; m sec -1 ) and SP (green contours, mbar)}
\end{figure}

\begin{figure}
        \includegraphics[width=\textwidth]{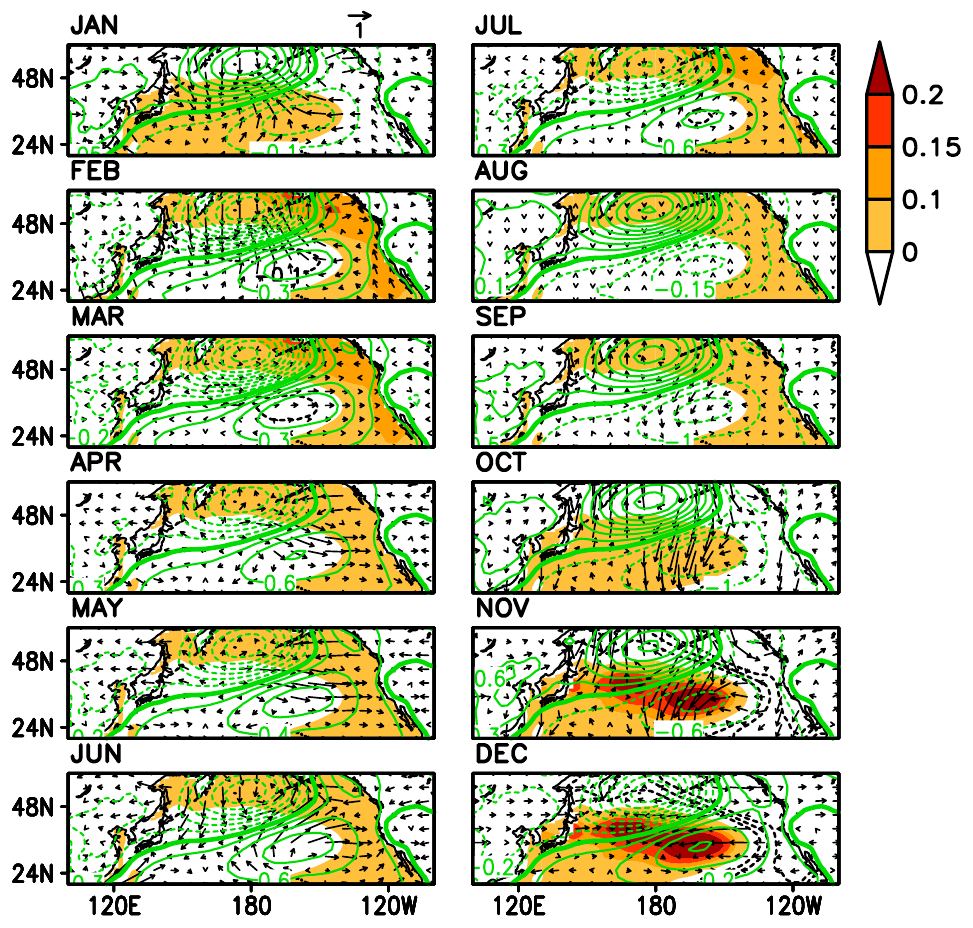}
        \renewcommand{\figurename}{Figure}
        \renewcommand{\thefigure}{S5}
        \caption{Long-term mean seasonal cycles of fourth SVD modes. SST (shaded; o C), wind (vectors; m sec -1 ) and SP (green contours, mbar)}
\end{figure}

\end{document}